\documentclass[10pt,twocolumn]{article}

\usepackage[utf8]{inputenc}
\usepackage[english]{babel}
\usepackage{amsmath}
\usepackage{amsfonts}
\usepackage{amssymb}
\usepackage{mathpazo} % Widely available alternative to Garamond
\usepackage[scaled]{helvet}
\usepackage[T1]{fontenc}
\usepackage{url}
\usepackage{verbatim}
\usepackage{float}
\usepackage{caption}
\usepackage{subcaption}
\usepackage{graphicx}
\usepackage{xcolor}
\usepackage{booktabs}
\usepackage{algorithm}
\usepackage[noend]{algpseudocode}
\usepackage{changepage}
\usepackage[normalem]{ulem}
\usepackage{soul}
\usepackage[left=48pt,%
                right=42pt,%
                top=42pt,%
                bottom=60pt,%
                headheight=15pt,%
                headsep=10pt,%
                letterpaper,twoside]{geometry}%
                
\usepackage[labelfont={bf,sf},%
                labelsep=period,%
                figurename=Fig.,%
                singlelinecheck=off,%
                justification=RaggedRight]{caption}
                
\usepackage[numbers,sort&compress]{natbib}
\title {10$^4$-
fold amplification of a tiny magnetic field to megagauss scale in femtosecond, ultraintense laser-solid interaction} 

\usepackage{authblk}
\author[1]{Anandam Choudhary}
\author[2]{Trishul Dhalia}
\author[1]{Sagar Dam}
\author[1]{Ameya Parab}
\author[1]{SK Rakeeb}
\author[1]{C Aparajit}
\author[1]{Amit D Lad}
\author[1]{Yash M Ved}
\author[3,4]{Kandaswamy Subramanian}
\author[2,*]{Amita Das}
\author[1,*]{G.Ravindra Kumar}

\affil[1]{Tata Institute of Fundamental Research, 1 Homi Bhabha Road, Colaba, Mumbai 400 005, India}
\affil[2]{Indian Institute of Technology, Hauz Khas, New Delhi 110016 }
\affil[3]{Ashoka University, Sonepat, Haryana 131029}
\affil[4]{IUCAA, Post Bag 4, Ganeshkhind, Pune 411007}

\affil[*]{Corresponding author: amita@iitd.ac.in; and grk@tifr.res.in}

\date{}

\begin{document}

\twocolumn[
  \begin{@twocolumnfalse} 
    
    \maketitle
    
    \begin{abstract}

Generating a powerful and quasistatic magnetic field within the confines of a tabletop laboratory experiment has proven to be a persistent challenge. The creation of magnetized high energy density plasma through such experiments presents significant opportunities for exploring several terrestrial  as well as astrophysical phenomena, apart from  controlling relativistic electron transport, directly relevant `for fusion schemes. 
Here  we demonstrate that the modest magnetic field (10$^{-3}$ megagauss ) in a common, readily available Neodymium magnet is amplified to  10's of megagauss levels lasting a few picoseconds, when excited by an ultraintense, femtosecond laser pulse. The experimental findings are strongly supported by particle-in-cell simulations, which not only validate the observations but also unveil a potential dynamo mechanism responsible for the enhancement and amplification of the axial magnetic field. These outcomes are of utmost importance in comprehending the intricacies of relativistic electron transport and the realm of magnetized laboratory astrophysics.

    \vspace{2mm}
    \end{abstract}

  \end{@twocolumnfalse}
]
\begin{figure*}[!ht]
   \centering
    \includegraphics[width=0.85\linewidth]{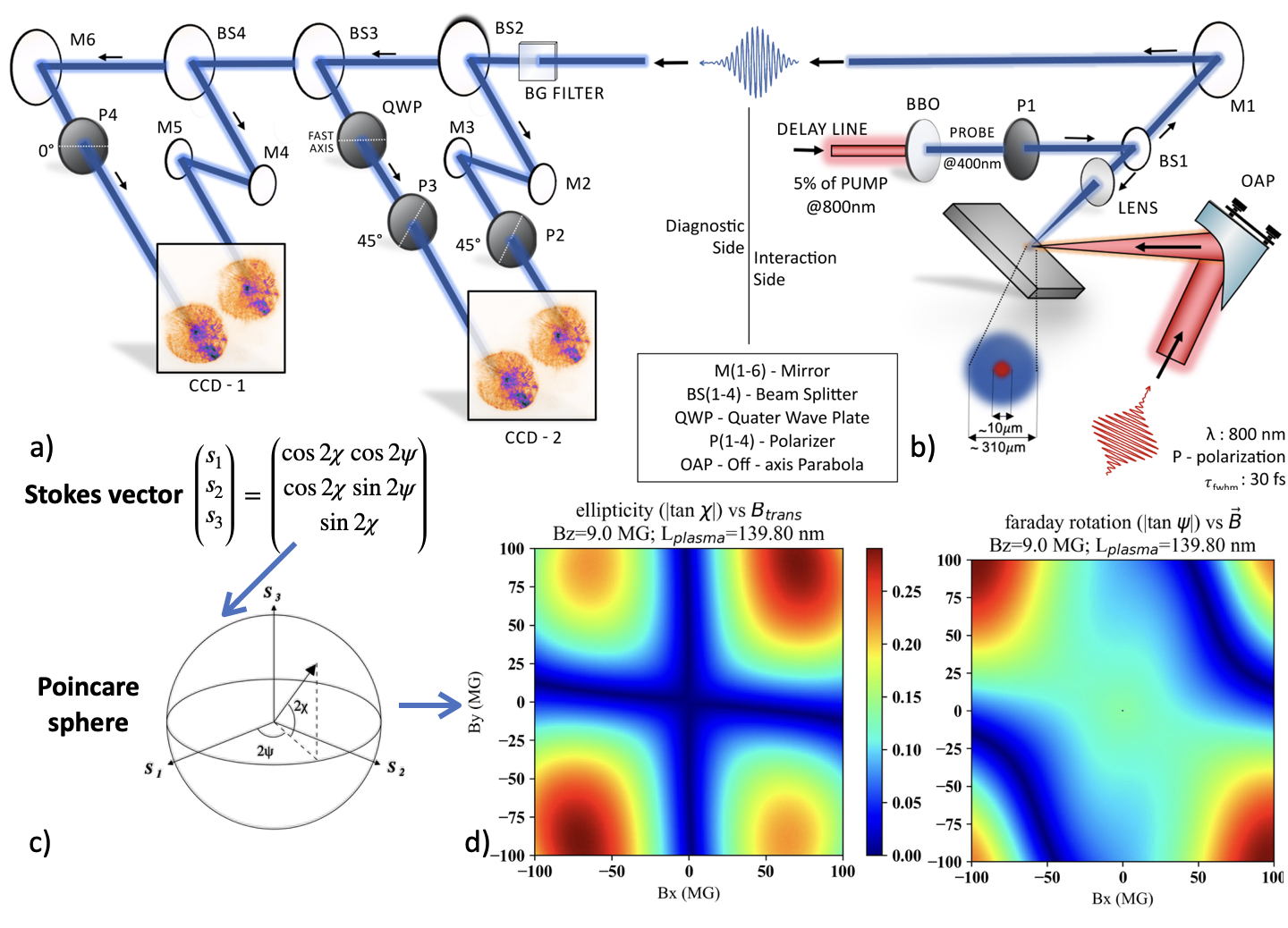}
   \caption{Experimental setup and magnetic field estimation process flow. (a) shows the polarimetry setup to measure all Stokes' components in a single laser shot. The reflected 400nm probe splits into 4 parts and passes through combinations of polarizers and wave plates. (b) main laser pulse irradiates the target and the delayed probe traverses and reflects inside the laser generated plasma. (c) signals from CCDs give stokes' vectors values and then ellipticity and Faraday rotation values are estimated. (d) Numerical integration is done to generate mappings between magnetic fields and observed ellipticity and Faraday rotation.}
   \label{fig:exp_setup}
\end{figure*}

\section{Introduction}
Magnetic fields pervade the universe, ranging from the microgauss scale in interstellar medium to trillions of Gauss in a neutron star \cite{beck2019synthesizing, han2002milestones}.  The origins and the evolution in magnitude of these fields in the cosmos are matters of great debate till date \cite{durrer2013cosmological, subramanian2016origin, shukurov2021astrophysical, kulsrud2008origin}. Terrestrially, the highest macroscopic scale magnetic fields are either generated by giant pulsed currents in normal \cite{nakamura2018record} or superconducting coils \cite{zhu2023china}. Laser produced high density plasmas on the other hand, hold the record for the largest (nearly billion gauss) fields measured on the earth, albeit on a microscopic length scale and nanosecond and picosecond time scales \cite{tatarakis2002measuring, wagner2004laboratory, 2020-4-Amita-RMPP-summary, sandhu2002laser, mondal2012direct, shaikh2016megagauss, chatterjee2017magnetic}.  The urge for  larger fields is usually met by flux compression \cite{jiang2021magnetic, GotchevPhysRevLett.103.215004} though it has its limitations. On the cosmic scale,  magnetic field amplification is typically explained by the dynamo mechanism \cite{subramanian2019primordial, brandenburg2005astrophysical, wagstaff2014magnetic, bhat2016unified}, where a miniscule field  caused by fluctuations in  astrophysical plasma are amplified   to large currents in a liquid/gaseous medium. There  have been some large scale experiments to simulate the dynamo in the laboratory in the steady state that invoke fluid instabilities in rotating conducting liquids \cite{gailitis2002colloquium, gregori2015generation} and laser plasma experiments where   plasma turbulence plays a role \cite{bott2021time, jiang2021magnetic}.   

  Given the fundamental interest in  the nature and magnitudes of  cosmic and terrestrial magnetic fields it is but natural that as much information about them is gathered under well controlled, easily reproducible, reasonably modest experimental settings. This is precisely what we address in this paper. We choose the contemporary and highly exciting setting of an ultrashort, high intensity laser pulse induced high energy density plasma in a solid on a 
 modest size laboratory tabletop and investigate  how a small, static magnetic field aids the generation of a giant magnetic field. We compare this with a magnetic field measurement in  the absence of the  static magnetic field and clearly demonstrate the dynamo effect. We perform particle-in-cell simulations that elucidate and enlarge the scope of the measurements and the derived physics.  An additional and important implication of our study is for the problem of relativistic electron transport in dense media, of great interest for femtosecond duration, hard EM radiation and  energetic particle sources and laser fusion.

 Our experiments are performed in the pump-probe configuration where a femtosecond, ultrahigh intensity laser pulse creates a high energy density plasma and a much weaker, time delayed probe femtosecond pulse, gathers information in the plasma  via  the magneto-optic Cotton-Mouton effect. The laser pulse conditions (polarization and angle of incidence  are chosen such that predominantly azimuthal fields are created by  the relativistic 'hot'  electrons streaming  into the target along the axial (normal to the planar target) direction.  
 The main experimental target chosen is a 5 mm thick, optically polished neodymium magnet with an inherent axial magnetic field of 1 kilo Gauss (0.1 T).  The magnetic fields generated in the laser-produced high energy density plasma are measured via the Stokes vectors measurement of the plasma interacted probe, and the fields' evolution is captured up to picoseconds. An unmagnetized neodymium target of the same geometry is used as a reference for comparing the generated magnetic fields. To understand the data and unravel the physics, we perform 2D OSIRIS particle-in-cell simulations \cite{hemker2000particle,fonseca2002osiris}. Both the experiments and simulation reveal a giant amplification of the initial, inherent magnetic field indicating a dynamo mechanism for the observed giant axial field. These giant magnetic fields of $\sim 10\: MG$ range can help to probe the regime of magnetized plasmas $(\omega_{ce}\sim \omega_{pe})$. Recent simulations have predicted potential applications of magnetized plasma in many areas of the interest such as high harmonic generation, parametric processes and laser energy absorption   \cite{dhalia2023harmonic,PhysRevE.110.065213,vashistha2021excitation,juneja2024enhanced,Juneja_2023,goswami2022observations}.   
 
 \section{Experiment}
The experiment was conducted with the Tata Institute of Fundamental Research 150 TW laser system. As shown in the Figure \ref{fig:exp_setup}, 5mm thick neodymium targets (magnetized and un-magnetized) were irradiated by 800 nm, 30 femtosecond p-polarised laser pulses focused to a $10\:\mu m$ spot (FWHM) by an $f/3$ off-axis parabolic mirror at $45^{\circ}$ incidence angle. Peak intensity of $\sim8\times$ 10$^{18}$ W/cm$^{2}$ was achieved with $10^{-7}$ intensity contrast at 20 picoseconds. The strength of the magnetic field in the magnetized target was 0.1 tesla directed normal to the surface.

The probe pulse, whose frequency is doubled (Figure \ref{fig:exp_setup} ) and effectively attenuated to a lower intensity (less than $10^{11}$ W/cm$^{2}$), is derived from the primary interaction pump through a thin beam-splitter. Second harmonic generation is accomplished employing a Type I crystal of SHG $\beta$-barium borate (BBO). The temporal delay of the probe pulse is achieved by utilizing a retro-reflector mirror mounted on a controlled delay stage (Physik Instrument). Our temporal measurement resolution is confined to the width of the probe pulse ($\sim$ 50 fs). Additionally, the second harmonic probe can penetrate the overdense plasma up to 4$n_c$ (where $n_c$ represents the critical density of the primary interaction pulse).

Another notable advantage of employing the second harmonic probe is eliminating noise originating from pump irradiation when the detector is coupled with a BG-39 filter. The probe is precisely focused perpendicular to the target, aligning with the direction of the density gradient. This specific orientation is crucial to prevent polarization-dependent phase shifts in the probe that may occur if the propagation direction deviates from parallel to the density gradient. Such deviations could introduce errors in magnetic field measurements, as changes in polarization are solely attributed to the presence of a magnetic field in the plasma. The probe focal spot size is approximately 300 $\mu m$, capturing the entire laser-excited area.
Before the experiment, the spatial overlap and temporal synchronization of the pump and probe were meticulously ensured. The temporal concurrence of the pump and probe pulses, as depicted in the figure \ref{fig:exp_setup}, is determined by a sharp transition in the reflectivity and transmissivity of the probe from the laser-created plasma. The experiment was conducted within a vacuum chamber under a pressure of $10^{-5}$ Torr.

\subsection*{Polarimetry and Data analysis}
The complete polarization state of the reflected probe is determined based on the experimentally measured Stokes' vectors. When a laser-produced plasma is subjected to a transient magnetic field, it behaves as a birefringent medium, causing the probe pulse to split into ordinary and extraordinary modes. These modes acquire different phases, resulting in the reflected probe attaining general polarization states with finite ellipticity (Cotton-Mouton effect) and orientation angle rotation (Faraday rotation effect) in the presence of transverse and axial magnetic fields, respectively.

The schematic of the Stokes' vector measurement setup is illustrated in the figure \ref{fig:exp_setup}. The reflected probe is divided into four parts using beam splitters and analyzed using a high-optical-quality Glan-air polarizer (Leysop) with a high extinction ratio of $10^{-6}$. The signal in the detector is obtained in terms of Stokes vectors in the following order: $I_D = S_0$ (direct reflectivity), $I_{0^0} = \frac{1}{2}(S_0+S_1)$, $I_{\lambda/4,45^0} = \frac{1}{2}(S_0-S_3)$, and $I_{45^0} = \frac{1}{2}(S_0+S_2)$, where $S_0$, $S_1$, $S_2$, and $S_3$ are the components of the Stokes vector. Detailed discussions on Stokes' vector and polarimetry can be found in \cite{segre1999review}. 

Four unknowns are solved using four equations to obtain all four Stokes' components. The ellipticity ($\chi$) and Faraday rotation ($\psi$) values can be directly deduced from the measured Stokes' vectors using the formulas \cite{segre1999review}:
$$\chi = \frac{1}{2}tan^{-1}\left(\frac{s_3}{\sqrt{s_1^2+s_2^2}}\right)$$ 
$$\psi = \frac{1}{2}tan^{-1}\left( \frac{s_2}{s_1}\right)$$

Here $s_1$, $s_2$ and $s_3$ are reduced Stokes' vector components (Stokes vector normalized with $S_0$). This polarization data was captured for different delays.

\smallskip
\textbf{Estimation of magnetic fields:} As the probe travels through the plasma up to its critical layer, it encounters both axial and transverse magnetic fields, leading to changes in the Stokes' vector (polarization states of the reflected probe pulse). The transverse field induces ellipticity gain (Cotton-Mouton effect) in the traversing probe pulse, and the axial field causes Faraday rotation of the probe's polarisation orientation. This way, in the presence of magnetic fields, the Stokes' vector evolves, and it is best described by Stokes' vector evolution equation \cite{segre1999review}. Numerical integration of the Stokes' vector evolution equation over the plasma slab scale length relates the changes in Stokes' vector with the magnetic fields. So with the experimental measured Stokes' components the magnetic field values are calculated. Due to few noise sources (fluctuations in laser parameters, finite effective extinction of polarizers), our magnetic field measurement resolution is limited to 1 Mega Gauss. This mapping enables the determination of the best-fit values of magnetic fields corresponding to the experimentally measured Stokes' vector changes at each delay. Both transverse- and axial-measured magnetic fields are averaged over the plasma scale length and transverse span.\\

To estimate the plasma scale length $L$ at each delay, we have employed the self-similar free expansion model \cite{wilks1997absorption}. This model assumes the isothermal expansion of plasma in a vacuum quantified using two fluid equations in one dimension. The density scale length relation from this model is $L=c_st$ ($c_s$ ion acoustic speed; $t$ probe delay with pump). The ion acoustic speed ($c_s$) of the density layer is estimated from our separate measurements of Doppler shifts and this is measured to be in the range $10^6-10^7\:\:cm/s$ \cite{jana2018probing, shaikh2018tracking}.\\

\begin{figure*}[!ht]
   \centering
    \includegraphics[width=1.0\linewidth]{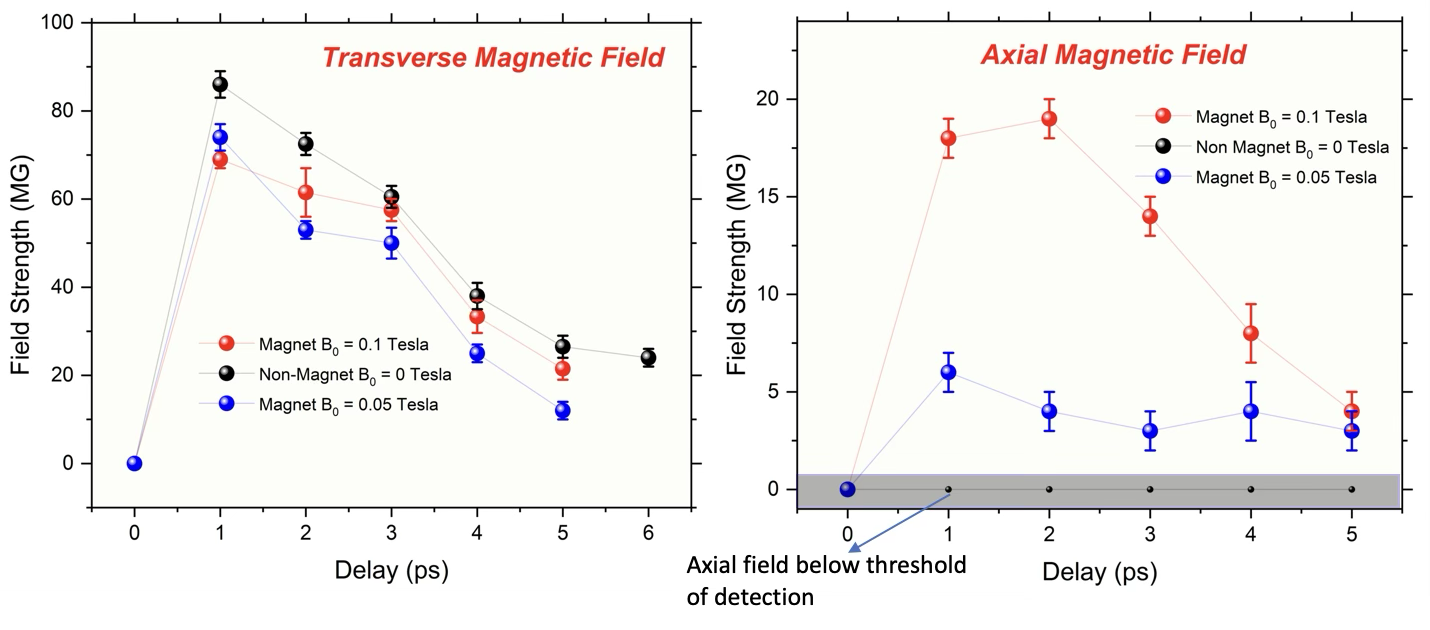}
   \caption{Magnetic field temporal growth shown for both magnetic and non-magnetic targets. An unmagnetized case is shown with zero initial field ($B_\circ=0$). Magnets with two different strengths (0.1 and 0.05 Tesla) are employed in the experiment.}
   \label{fig:mag_field_exp}
\end{figure*}
\begin{figure}[!ht]
   \centering
    \includegraphics[width=1\linewidth]{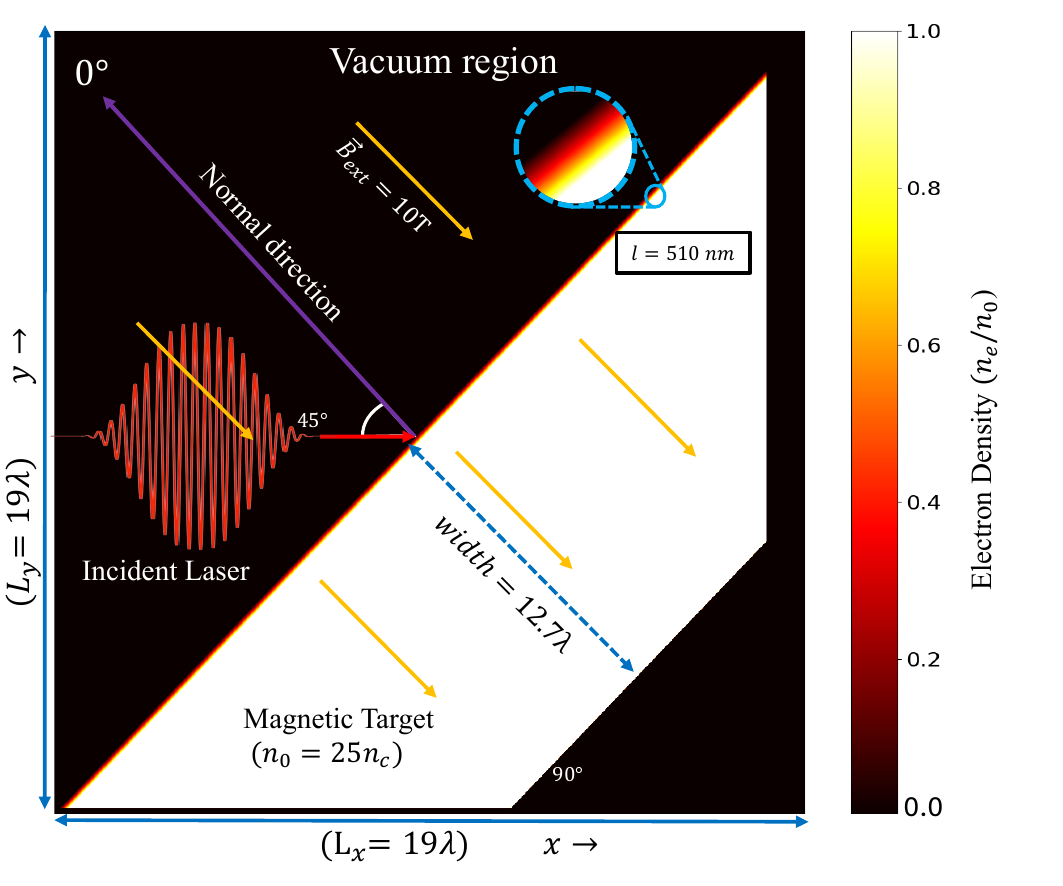}
   \caption{Schematic of simulation geometry}
   \label{fig:schematic_dynamo}
\end{figure}

\begin{figure*}[!ht]
   \centering    \includegraphics[width=0.9\linewidth]{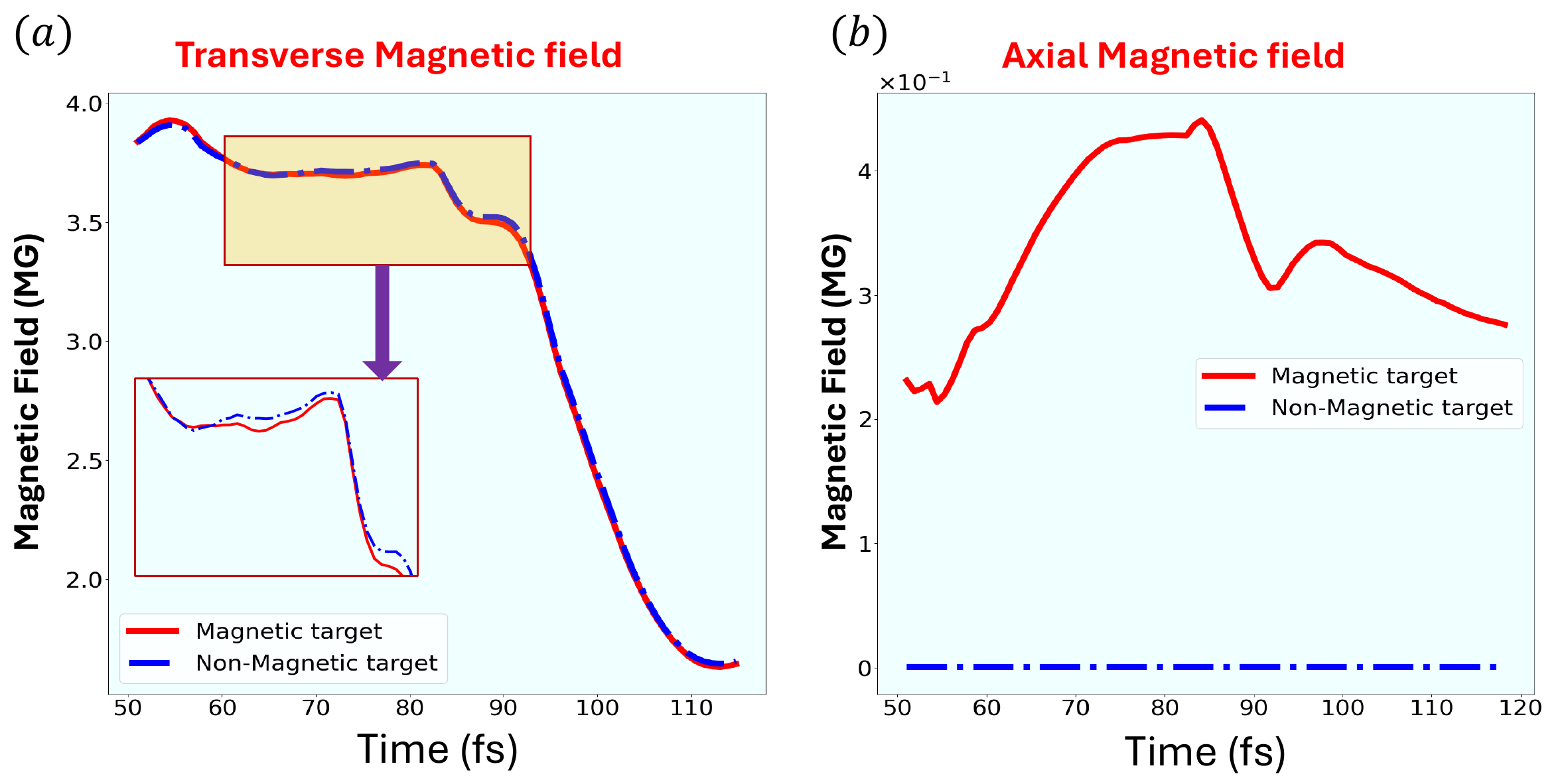}
   \caption{Magnetic field time evolution for both magnetic (10 T) and non-magnetic (0 T) target from simulation.   }
   
   \label{fig:mag_field_simulation}
\end{figure*}

\begin{figure*}[!ht]
   \centering
    \includegraphics[width=0.8\linewidth]{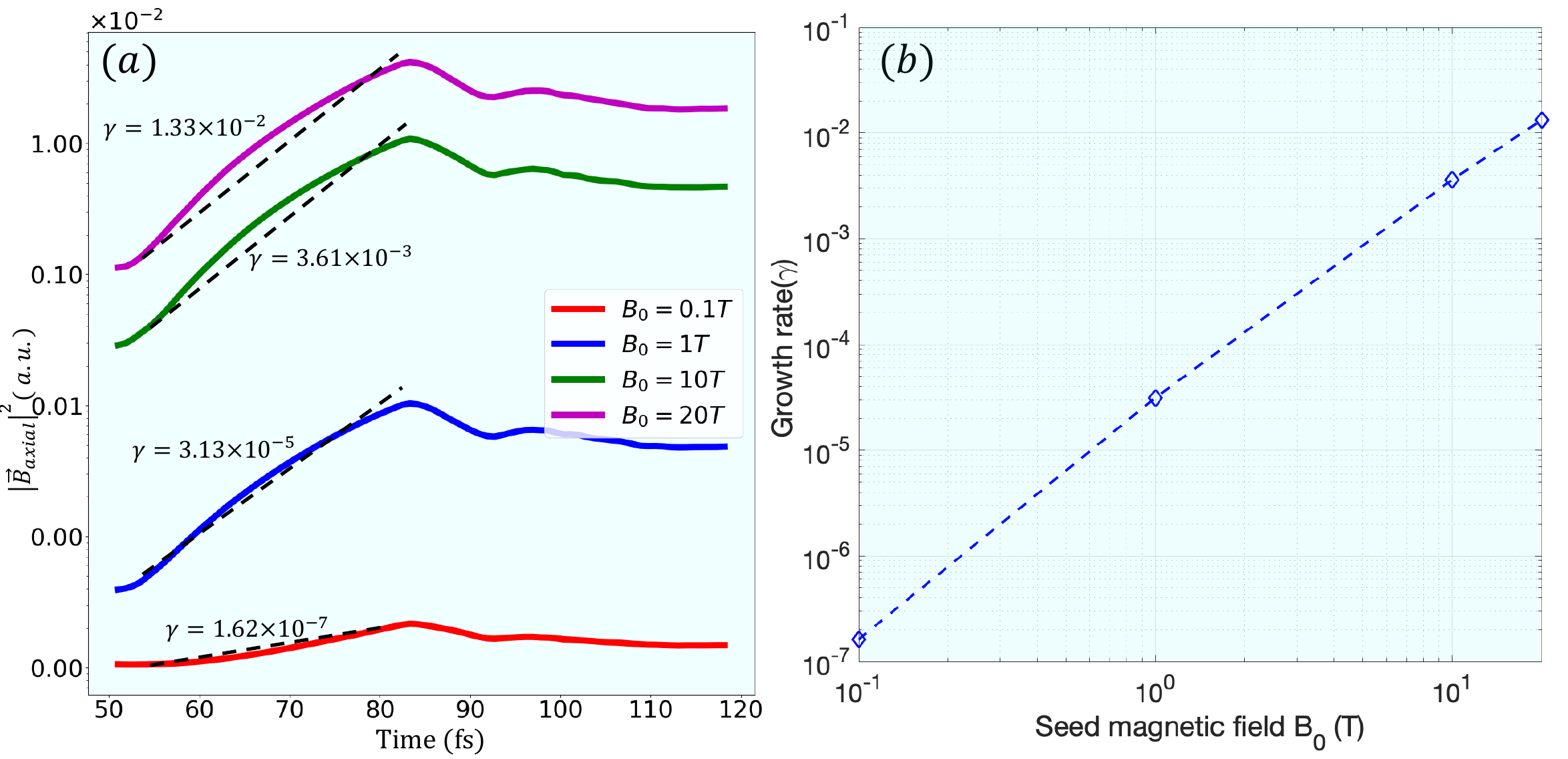}
   \caption{Figure (a) presents time evolution of axial magnetic field energy (on log axis)  for different values of seed magnetic field. Slope (growth rate) of these curves with respect to applied seed magnetic field has been plotted in figure (b).}
   \label{fig:growth_rate_complete}
\end{figure*}
The striking observation from the experiments is shown in Figure \ref{fig:mag_field_exp}, which shows the evolution of the mean transverse and axial magnetic fields over picosecond timescales. The mean axial field is significantly amplified for magnetic targets ($B_0 = 0.1\: T,\: 0.05\: T$) in contrast to non-magnetic targets ($B_0 = 0\: T$). The transverse field evolves similarly in all cases, reaching up to 86 MG for the non-magnetic target and up to 70 MG for the magnetic targets. In contrast, the axial fields are completely absent (i.e., below the detection threshold) for the non-magnetic target which shows a prominent four-order-of-magnitude amplification in the presence of even a small initial magnetic field (0.1 T). These fields subsequently decay over a period of up to 6 picoseconds. The amplification of seed static magnetic fields to megagauss levels motivated us to further investigate the underlying processes through simulations, as experimental measurements alone cannot resolve the fine temporal dynamics that may lead to a magnetic dynamo. The next section describes the Particle-in-Cell (PIC) simulations we performed to model our experiment, providing a comprehensive interpretation of our experimental results.
\begin{figure*}[!ht]
   \centering
    \includegraphics[width=0.9\linewidth]{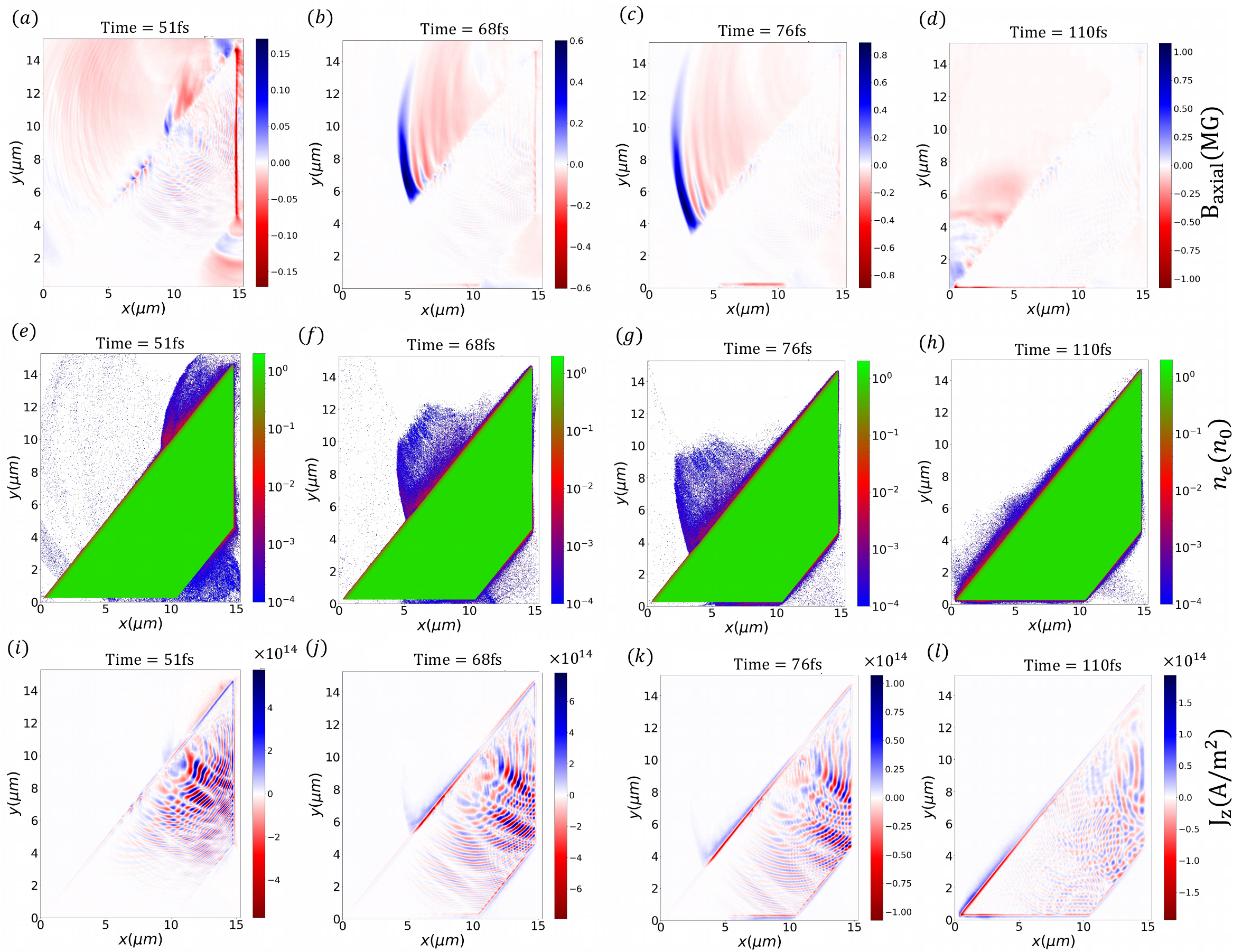}
   \caption{Upper panel shows snapshot of axial magnetic field ($|B_xcos45 -B_ysin45|$) for magnetic ($10t$) target at various times and middle panel shows snapshots of electron density at various times. The lower panel presents snapshots for the z-direction current density component ($J_z$) at various times.}
   \label{fig:axial_current_density}
\end{figure*}

\begin{figure}[!ht]
   \centering
    \includegraphics[width=0.85\linewidth]{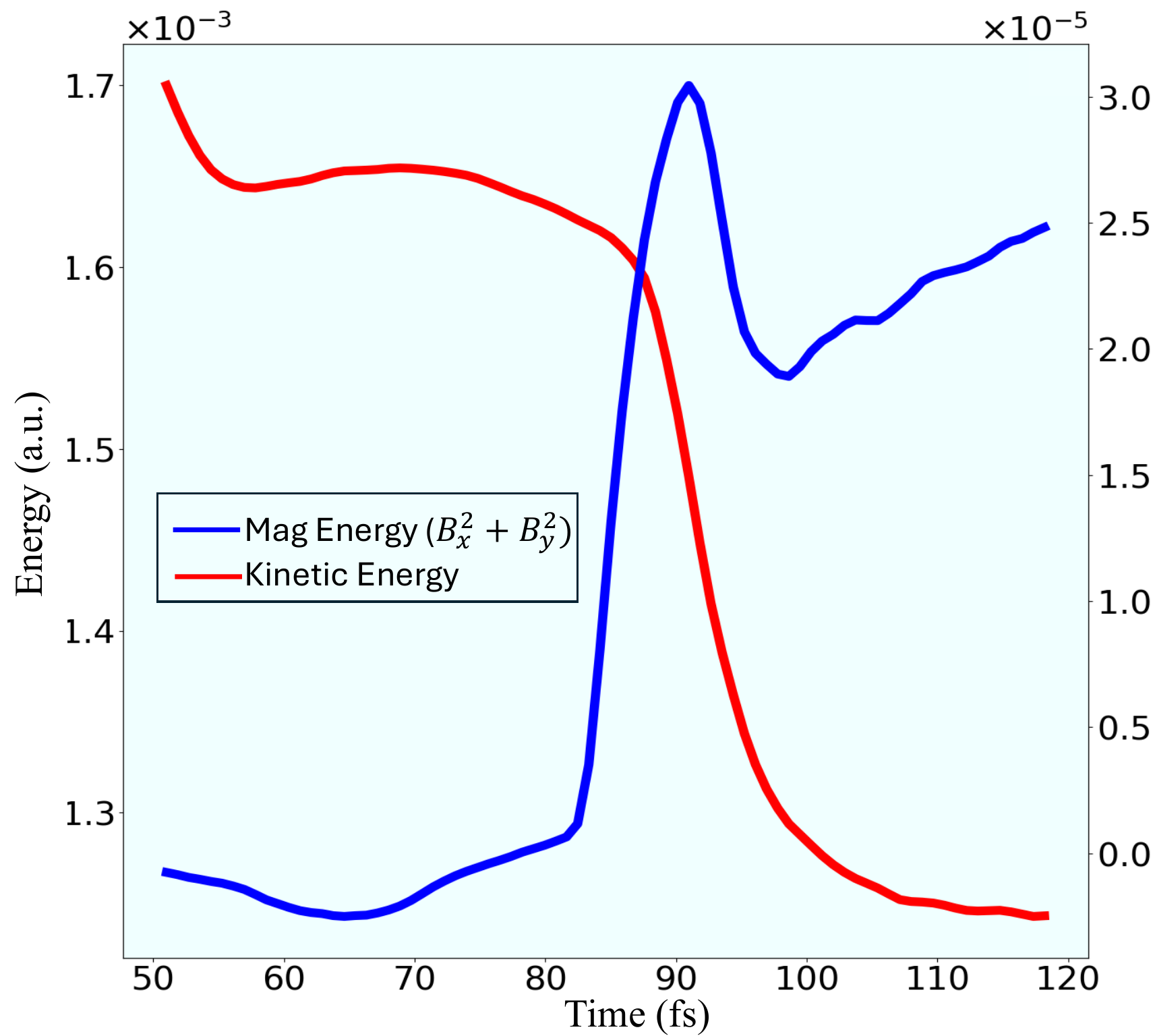}
   \caption{ Time evolution of Kinetic energy of electrons and magnetic energy in plane for the magnetic target $(B_0=10T)$.   }
   \label{fig:energy_evolution}
\end{figure}

\begin{figure*}[!ht]
   \centering
    \includegraphics[width=0.75\linewidth]{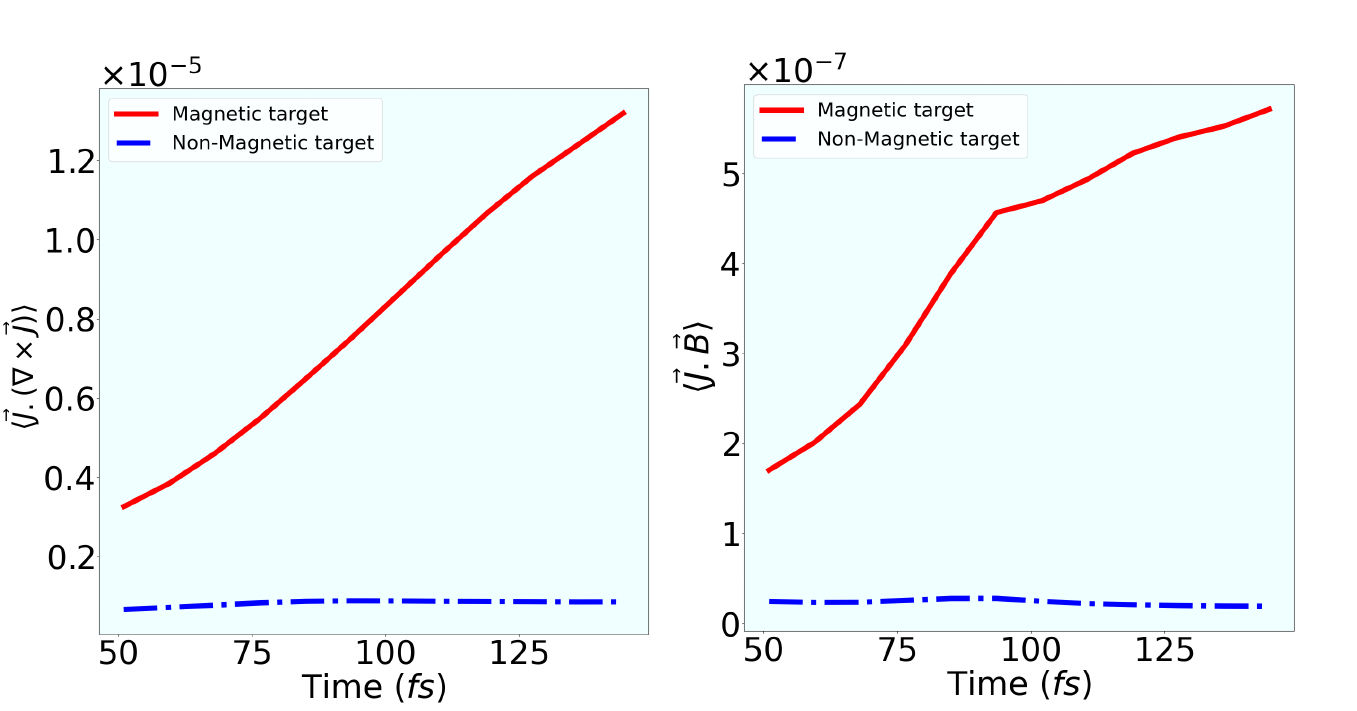}
   \caption{Helicity for both magnetic ($10T$) and non-magnetic target with time.   }
   \label{fig:helicity}
\end{figure*}
\section{Simulation and Observations}
PIC simulations to understand the process of the observed amplification of the seed magnetic field were conducted using OSIRIS 4.0 \cite{hemker2000particle,fonseca2002osiris}. It should be recognized that the experimental measurements have been carried out over a long duration ($\sim picoseconds$) and for a large system size of ($\sim cms$) scales  \cite{choudhary2023controlling}. The exact replication of the experimental situation is, therefore, not possible in simulations. Even with significantly advanced computational resources, it would still be challenging to precisely replicate the experiments. Our aim here is to explore a similar regime of physics, albeit operative at shorter  times and smaller regime. Schematic of simulation geometry chosen has been presented in figure \ref{fig:schematic_dynamo}. We chose a 2-D  simulation box of size $19\lambda \times 19\lambda$ considerably small compared to the size of the plasma created in the experiment. The plasma density is also chosen to be smaller here $25 n_c$ compared to the experimental value of $100n_c$ (here $n_c$ is the critical density of the plasma for the incident laser). Thus, in simulation electron skin depth is longer for which a coarser grid size suffices. To mimic a high Z material,  a fully ionized plasma with a neutralizing (initially) immobile ion background has been considered. An exponential density gradient of the form $n=25n_c(\exp((x-y)\ln(2)/L)-1)$ at the front surface has been considered with a preplasma scale length of $L=510 nm$ and longitudinal width of $12.7 \lambda$.
A linear p-polarized laser pulse with wavelength $\lambda=800 nm$ is incident at AOI $45^\circ$ on the plasma surface. The normalized laser vector potential has been chosen ($a_0$=$eE/m_e\omega_lc$=$6.1$) corresponding to $I$ =$5.65\times10^{19}W/cm^2$. The laser pulse duration of $5$ laser cycles has been considered. The pulse width of the incoming wave pulse is taken $25.5 fs$ with a spot size of $3.56 \mu m$. These values are typically comparable with experiments except that the laser spot size in experiments is about  $10\: \mu m$. 

The typical time scales both in experiments and simulations are slower than the plasma frequency $\omega_{pe}$. The experimental observation time $ t_{obs} \sim \omega_{obs}^{-1}\sim ps$ is much shorter than the electron gyro-period $\omega_{ce}^{-1} \sim ns$. However, it should be noted that the condition of $\omega_{obs} < \omega_{pe}^2/\omega_{ce}$ is satisfied. This suggests that the dynamics of the electron species after the laser leaves the system can essentially be looked upon to be governed by the Electron Magnetohydrodynamics (EMHD)  fluid depiction \cite{das1999nonlinear,yadav2008propagation,das2000theory}. Electron Magnetohydrodynamics (EMHD) is a reduced fluid depiction of plasma. However, unlike Magnetohydrodynamics its applicability is at fast time scales for which the dynamical response of ions is irrelevant. It depicts phenomena concerning strong currents associated with electron flow. Furthermore,  here too like MHD the displacement current in Maxwell's equation has a negligible role to play. This implies that   $\omega < \omega_{pe}^2/\omega_{ce}$ (where $\omega^{-1}$ is typical time scale of the problem. While MHD depiction has provided many insights in the context of a variety of plasmas such as solar, stellar, astrophysical, and magnetic fusion, where time scales are long and ion dynamics in important, EMHD has traditionally been employed in the context of a variety of applications such as fast switches \cite{jain2003nonlinear,bulanov1992magnetic}. It is interesting to note that it is also applicable to depict the dynamics immediately after the laser pulse has interacted with the plasma medium. This is indeed borne out by the time scales that we have outlined above.  

The hurdle that one wishes to now overcome is the fact that the typical growth of the magnetic field in experiments occurs at picosecond time scales. Simulations involving the initial phase of laser interaction with plasma cannot be continued till this time scale. We, therefore, need to somehow hasten this magnetic field amplification process to occur at around  $ \sim 100 fs$. Also, one has the constraint of having the phenomena show up within the smaller spatial regime of the simulation box. Keeping this in view, we chose an enhanced value for the initial applied magnetic field $10 \ Tesla$ in our simulation. It should be noted that both experiments and simulations provide evidence of faster growth of the axial magnetic field with increasing value of the applied magnetic field. Figure \ref{fig:mag_field_exp}(b) has shown that in experiments for different applied seed magnetic fields (0.05 T and 0.1 T), amplification of the axial magnetic field is enhanced when the seed field is increased. In simulation we will also see a similar amplification growth with increasing seed field. Figure \ref{fig:mag_field_simulation} (a) and (b) depict the time evaluation of the generated mean transverse and axial magnetic field respectively for both magnetic and non-magnetic targets. At about  $80fs$ a four-fold amplification $(40T)$ in mean axial magnetic field is observed. The subsequent sharp decline essentially happens as the result of electron currents leaving the simulation box. In experiments also this decline occurs, which could be due to recombination etc., leading to diminishing  plasma density. Also, in figure \ref{fig:growth_rate_complete}(a) for different applied initial seed magnetic fields, the growth rate of axial magnetic field also increases in simulations. The growth rate of axial magnetic field has been calculated as, 
$$    \gamma= \frac{1}{2t}log_e \left|\frac{B_{axial}}{B_0}\right|^2$$

Figure \ref{fig:growth_rate_complete} (b) shows that growth rate of axial magnetic field increases almost in linear fashion against applied seed magnetic field. Thus the choice of a higher value of the magnetic field aligns with our aim of having a qualitative comparison with the experimental result through simulations and then identifying a detailed understanding of the underlying process. 
The features of experimental observation are also observed in our simulation where we observe amplification of axial magnetic field for magnetic targets.  

Let us now try to understand the process involved in this enhancement of the the axial magnetic field in detail. The laser pulse stirs the lighter electron fluid which gains kinetic energy and also extracts them out from the target surface as seen in figure \ref{fig:axial_current_density} second row. The electron bunch created on the target surface tends to travel along the target surface. This propagation is similar for both magnetic and non-magnetic targets. This clearly shows that the external applied magnetic field is far too weak to significantly influence the electron dynamics.  This is also borne out from our earlier evaluation of time scales showing that the observation time does not even correspond to a single electron gyro period. 

A comparison of the second and first row, however, shows that as this electron bunch propagates there is a strong magnetic field generation at its edge. This can be understood from EMHD fluid depiction in which  the field $\vec{B} - d_e^2 \nabla^2 \vec{B}$ is supposedly tied. The propagation thus leads to the magnetic field lines getting dragged and condensing at the edge. Furthermore, the conversion of electron kinetic energy to magnetic field energy may also happen in the case of EMHD. This is also observed from Fig \ref{fig:energy_evolution} where a concomitant increase in magnetic field energy with decreasing kinetic energy of the electron is observed.

\section{Discussion and conclusion }
We have observed the amplification of axial magnetic fields for magnetic targets in both experiments and simulations. The non-magnetic targets do not exhibit such a feature. The theory of magnetic field amplification in the context of MHD has been studied extensively and the dynamo mechanism is often invoked in MHD for the same. Here, the plasma dynamics is essentially dictated by  EMHD fluid - which has significant differences with MHD. Furthermore, the understanding of fundamental processes leading to magnetic field amplification, diffusion etc., are  fairly well understood in the context of MHD. A dynamo mechanism is often invoked for the amplification purpose which requires helical flows having  $\alpha \sim \int \vec{v} \cdot \nabla \times \vec{v} \hspace{0.1in} d^3r$ as finite. 

We also seek the possibility of the connection of amplification of magnetic field with helical flows in this case of our for the flow of  electron fluid. The plot of average value of $\vec{J} \cdot \nabla \times \vec{J} $ from figure \ref{fig:helicity}$(a)$ in a sense represents that the helicity shows a steady increase with time. We have also plotted $\vec{J} \cdot \vec{B} $ in figure \ref{fig:helicity}$(b)$ which also shows a steady increase.

In the context of EMHD the understanding of the processes related to amplification and diffusion of magnetic field is still at a quite nascent stage. 
We have in this work provided an experimental platform where amplification has been observed. The PIC simulation also provides support for the same.\\ 
 
%\section{Conclusions}

% \cite{Remingtondoi:10.1126/science.284.5419.1488}
% \cite{Hayes2020}
%\cite{Purvis2013}
%\cite{Curtis2018}
%\cite{Bargstendoi:10.1126/sciadv.1601558}
%\cite{Pile2021}
%\cite{Kokurewicz2021}\\

\noindent\begin{Large}

\textbf{Acknowledgement}\end{Large}
 GRK acknowledges partial support from J.C. Bose Fellowship grant (JBR/2020/000039) from the Anusandhan National Research Foundation (ANRF), Government of India. AD acknowledges support from  the ANRF core grants CRG 2018/
 000624 and CRG/2022/002782 as well as a J C Bose Fellowship grant JCB/2017/000055. AD and
TD would like to acknowledge the OSIRIS Consortium,
consisting of UCLA and IST (Lisbon, Portugal) for
providing access to the OSIRIS framework which is
the work supported by NSF ACI-1339893. TD would like to thank IIT Delhi HPC facility for computational resources and the Council for Scientific and Industrial Research (Grant No-09/086/(1489)/2021-EMR-I) for funding the research. ADL acknowledges support from the Infosys-TIFR Leading Edge Research Grant (Cycle 2).
% Bibliography
\bibliography{references}
\bibliographystyle{ieeetr}

\end{document}